\documentstyle[12pt,aps,psfig]{revtex}
\newcommand{\be}{\begin{equation}}
\newcommand{\ee}{\end{equation}}
\newcommand{\bel}[1]{\be\label{#1}}
\newcommand{\re}[1]{Eq.~(\ref{#1})}
\newcommand{\mbs}[1]{\mbox{$\scriptstyle{#1}$}}
\newcommand{\ds}{\displaystyle}
\newcommand{\qq}{\mbox{$q \overline{q}$}\,\,}
\newcommand{\mub}{\overline{\mu}}
\newcommand{\psib}{\overline{\psi}}
\newcommand{\rhob}{\overline{\rho}}
\newcommand{\bm}[1]{\mbox{\boldmath${#1}$\unboldmath}}

\newcommand{\intp}{\int\frac{{\rm d}^3 p}{(2\pi)^3}}
\newcommand{\ocnf}{n_{\bm{p}f}}
\newcommand{\ocnbf}{\overline{n}_{\bm{p}f}}
\newcommand{\dd}{\partial\hspace{-6pt}/}
\newcommand{\dD}{D\hspace{-7pt}/}
\begin{document}
\renewcommand{\thefootnote}{\arabic{footnote}}
\begin{center}
{\Large\bf Strange quark matter within the Nambu--Jona-Lasinio model}\\[5mm]
{\bf I.N.~Mishustin$^{\,1,2,3}$,
L.M.~Satarov$^{\,1,2}$, H.~St\"ocker$^{\,2}$, and W.~Greiner$^{\,2}$}
\end{center}
\begin{tabbing}
\hspace*{1.5cm}\=${}^1$\,\={\it The Kurchatov~Institute,
Russian Research Centre, 123182~Moscow,~\mbox{Russia}}\\
\>${}^2$\>{\it Institut~f\"{u}r~Theoretische~Physik,
J.W.~Goethe~Universit\"{a}t,}\\
\>\>{\it D--60054~Frankfurt~am~Main,~\mbox{Germany}}\\
\>${}^3$\>{\it The Niels~Bohr~Institute,
DK--2100~Copenhagen {\O},~\mbox{Denmark}}\\
\end{tabbing}

\begin{abstract}

Equation of state of baryon rich quark matter is studied within the SU(3)
Nambu--Jona-Lasinio model with flavour mixing interaction. Possible
bound states (strangelets) and chiral phase transitions in this matter
are investigated at various values of strangeness fraction $r_s$. The
model predictions are very sensitive to the ratio of vector and
scalar coupling constants, $\xi=G_V/G_S$. At $\xi=0.5$
and zero temperature the maximum binding energy (about 15 MeV per
baryon) takes place at $r_s\simeq 0.4$.  Such strangelets are
negatively charged and have typical life times $\sim10^{-7}$ s. The
calculations are carried out also at finite temperatures. They show
that bound states exist up to temperatures of about 15 MeV. The model
predicts a first order chiral phase transition at finite baryon
densities. The parameters of this phase transition are calculated as a
function of $r_s$.

\end{abstract}

\baselineskip 24pt

\section{Introduction}

Almost 30 year ago A.B. Migdal has put forward a brilliant idea of
pion condensation in nuclear matter~\cite{Mig71}. Soon it was
realized that this phenomenon may lead to the existence of density
isomers~\cite{Mig77}. At about this time Lee and Wick proposed
another mechanism leading to the appearance of an abnormal nuclear
state~\cite{Lee74}. It is related to the restoration of chiral
symmetry at high baryon density. These ideas have motivated
nuclear community to initiate new experimental programs aimed at
producing hot and dense nuclear matter in energetic collisions of
heavy nuclei. These experiments have started in Dubna and Berkeley
and then continued in Brookhaven and CERN.  Nowadays exciting
expectations are associated with new ultrarelativistic heavy ion
colliders, RHIC and LHC.

The general goal of present and future experiments with
ultrarelativistic heavy ions is to study the equation of state and
dynamical properties of strongly interacting matter. Now the main
interest lies in investigating the chiral and deconfinement phase
transitions, predicted by QCD. The ultimate goal is to produce and study
in the laboratory a new state of matter, the Quark--Gluon Plasma (QGP).
This state of matter can be reached only at high temperatures or
particle densities when elementary constituents, quarks and gluons are
liberated from hadrons.

Since the direct application of QCD at moderate temperatures and
nonzero chemical potentials is not possible at present, more simple
effective models respecting some basic symmetry properties of QCD are
commonly used. One of the most popular models of this kind, which is
dealing with constituent quarks and respects chiral symmetry, is the
Nambu--Jona-Lasinio (NJL) model \cite{Nam61,Lar61}. In recent years
this model has been  widely used for describing hadron properties (see
reviews \cite{Vog91,Kle92}), phase transitions in dense matter
\cite{Asa89,Kli90a,Cug96,Kli97,Raj99,Kle99} and multiparticle bound
states \cite{Koc87,Bub96,Mis96,Bub99}.

In the previous papers \cite{Mis99,Mis00} we have used the NJL model to
study properties of the quark--antiquark plasma out of chemical
equilibrium. In fact, we considered a system with independent densities
of quarks and antiquarks. We have found not only first order
transitions but also deep bound states even in the baryon--free matter
with equal densities of quarks and antiquarks. In the present paper the
emphasis is put on investigating the possibility of bound states and
phase transitions in equilibrated matter at various flavour
compositions.  In particular, we consider the possibility of bound
states in quark matter with a significant admixture of strange quarks,
i.e. strangelets. Thermal properties of
strange quark matter are also studied.

The paper is organized as follows: in Sect.~II a generalized NJL model
including flavour--mixing terms is formulated in the mean--field
approximation. The model predictions for strange matter and
characteristics of its bound states at zero temperature are discussed
in Sect.~III. Effects of finite temperatures are considered in
Sect.~IV. Possible decay modes of new bound states are discussed in
Sect.~V. Main results of the present paper are summarized in Sect.~VI.

\section{Formulation of the model}

Below we use the SU(3)--flavour version of the NJL model suggested in
Ref.~\cite{Reh96}. The corresponding Lagrangian is written as
($\hbar=c=1$)
\begin{eqnarray}
{\cal L}=\psib\,(i\,\dd-\hat{m}_0)\,\psi&+&G_S\sum_{j=0}^{8}
\left[\left(\psib\,\,\frac{\ds\lambda_j}{\ds 2}\,\psi\right)^2+
\left(\psib\,\frac{\ds i\gamma_5\lambda_j}
{\ds 2}\,\psi\right)^2\right]\nonumber\\
&-&G_V\sum_{j=0}^{8}\left[
\left(\mbox{$\psib\,\gamma_\mu \frac{\ds\lambda_j}{\ds 2}\,\psi$}\right)^2+
\left(\mbox{$\psib\,\gamma_\mu \frac{\ds\gamma_5\lambda_j}
{\ds 2}\,\psi$}\right)^2\right]\nonumber\\
&-&K\left[{\rm det}_f\,\mbox{$\left(\psib\,(1-\gamma_5)\,\psi\right)$}+
{\rm det}_f\,\mbox{$\left(\psib\,(1+\gamma_5)\,\psi\right)$}\right].
\label{lagr}
\end{eqnarray}
Here~$\psi$~is~the~column~vector~consisting~of~three~single--flavour
spinors $\psi_f$, \mbox{$f=u,d,s$}, $\lambda_1,\ldots,\lambda_8$~~are
the SU(3) Gell-Mann matrices in flavour
space,~~$\lambda_0\equiv\sqrt{2/3}\,\bm{I}$, and
\mbox{$\hat{m}_0={\rm diag}(m_{0u},\,m_{0d},\,m_{0s})$}
is the matrix of bare (current) quark masses. At $\hat{m}_0=0$ this
Lagrangian is invariant with respect to
\mbox{${\rm SU_{\,L}(3)}\otimes\,{\rm SU_{\,R}(3)}$}
chiral transformations. The second and third terms in \re{lagr}
correspond, respectively, to the scalar--pseudoscalar and
vector--axial-vector 4--fermion interactions. The last 6--fermion
interaction term breaks the $U_A(1)$ symmetry and gives rise  to the
flavour mixing effects.

In the mean--field approximation the Lagrangian (\ref{lagr}) is reduced to
\begin{eqnarray}
{\cal L}_{\rm mfa}&=&\sum_f\psib_f\,(i\dD-m_f)\,\psi_f
-\frac{\ds G_S}{\ds 2}\sum_f\rho_{\mbs{Sf}}^2\nonumber\\
&+&\frac{\ds G_V}{\ds 2}\sum_f{\rho_{\mbs{Vf}}}^2 +
4K\prod_f\rho_{\mbs{Sf}}\,,\label{lagrm}
\end{eqnarray}
where $\dD=\dd+i\,\gamma_0 G_V\rho_{\mbs{Vf}}$ and
\begin{eqnarray}
\rho_{\mbs{Sf}}&=&<\psib_f\psi_f>\,,\label{densc}\\
\rho_{\mbs{Vf}}&=&<\psib_f\gamma_0\psi_f>\label{densv}
\end{eqnarray}
are scalar and vector densities of quarks with flavour $f$\,. Angular
brackets correspond to the quantum--statistical averaging. The
constituent quark masses, $m_f$, are determined by the coupled set of
gap equations
\bel{gape}
m_f=m_{0f}-G_S\,\rho_{\mbs{Sf}}+
2K\,\prod_{f'\neq f}\rho_{\mbs{Sf'}}\,.
\ee

The NJL model is an effective, non--renormalizable model.  To
regularize the divergent contribution of negative energy states of
the Dirac sea, one must introduce an ultraviolet
cut--off.  Following common practice, we use the 3--momentum
cut--off $\Theta(\Lambda-p)$ in divergent integrals
\footnote{Here $\theta\,(x)\equiv\frac{1}{2}\,(1+{\rm sgn}\,x)$.}.
The model parameters $m_{0f}, G_S, K, \Lambda$ can be fixed by
reproducing the observed masses of $\pi, K$\,, and $\eta'$ mesons as
well as the pion decay constant $f_\pi$. As shown in Ref.~\cite{Reh96},
a reasonable fit is achieved with the following values:
\begin{eqnarray}
m_{0u}=m_{0d}&=&5.5~{\rm MeV},~~~m_{0s}=140.7~{\rm
MeV},\\
G_S=20.23~{\rm GeV}^{-2}, &&K=155.9~{\rm GeV}^{-5},
~~~\Lambda=0.6023~{\rm GeV}.
\label{papa}
\end{eqnarray}
Motivated by the discussions in in Refs.~\cite{Vog91,Cas95}, we
choose the following value of the vector coupling
constant\footnote{See discussion of this question in Ref.~\cite{Mis00}}
\bel{coup}
G_V=0.5\,G_S=10.12\,{\rm GeV}^{-2}\,.
\ee

Let us consider homogeneous, thermally (but not, in general,
chemically) equilibrated quark--antiquark matter at temperature $T$\,.
Let $a_{\bm{p},\lambda}$\,(\,$b_{\bm{p},\lambda}$) and
$a^+_{\bm{p},\lambda}$\,(\,$b^+_{\bm{p},\lambda}$) be the destruction
and creation operators of a quark (an antiquark) in the state
$\bm{p},\lambda$\,, where $\bm{p}$ is the 3-momentum and $\lambda$ is
the discrete quantum number denoting spin and flavour (color indices
are suppressed). It can be shown~\cite{Mis99} that quark and antiquark
phase--space occupation numbers coincide with the Fermi--Dirac
distribution functions:
\begin{eqnarray}
<a^+_{\bm{p},\lambda}\,a_{\bm{p},\lambda}>\equiv\ocnf
&=&\left[\,\exp{\left(\frac{E_{\bm{p}f}-\mu_{Rf}}
{T}\right)}+1\,\right]^{-1}\,,\label{ocnq}\\
<b^+_{\bm{p},\lambda}\,b_{\bm{p},\lambda,f}>\equiv\ocnbf &=&
\left[\,\exp{\left(\frac{E_{\bm{p}f}-\mub_{Rf}}{T}\right)}
+1\,\right]^{-1}\,,\label{ocnqb}
\end{eqnarray}
where $E_{\bm{p}f}=\sqrt{m_f^2+\bm{p}^2}$ and $\mu_{Rf},\,\mub_{Rf}$
denote the reduced chemical potentials of quarks and antiquarks:
\begin{eqnarray}
\mu_{\mbs{Rf}}&=&\mu_f-G_V\rho_{\mbs{Vf}}\,,\label{rce1}\\
\mub_{\mbs{Rf}}&=&\mub_f+G_V\rho_{\mbs{Vf}}\,.\label{rce2}
\end{eqnarray}

The explicit expression for the vector density can be written as
\bel{vecd}
\rho_{\mbs{Vf}}=\rho_f-\rhob_f\,,
\ee
where
\bel{denf}
\rho_f=\nu\intp\,\ocnf,\,\,\,\,\rhob_f=\nu\intp\,\ocnbf\,
\ee
are, respectively, the number densities of quarks and antiquarks
of flavour $f$ and $\nu=2 N_c=6$ is the spin--color degeneracy factor.
The net baryon density is obviously defined as
\bel{bard}
\rho_B=\frac{1}{3}\sum_f\rho_{Vf}\,.
\ee
The physical vacuum ($\rho_f=\rhob_f=0$) corresponds
to the limit $\ocnf=\ocnbf=0$\,.

In general the chemical potentials $\mu_f$ and
$\mub_f$ are independent variables. The assumption of chemical
equilibrium with respect to creation and annihilation of \qq pairs
leads to the conditions
\bel{eqcp}
\mub_f=-\mu_f,~~f=u,d,s~.
\ee
These conditions automatically follow from the relations
valid for any \qq system chemically equilibrated
with respect to strong interactions:
\bel{ches}
\mu_i=B_i\,\mu_B+S_i\,\mu_S+Q_i\,\mu_Q\,\,\,(i=u,d,s,\bar{u},
\bar{d},\bar{s})\,.
\ee
Here three independent chemical potentials, $\mu_B,\,\mu_S,\,\mu_Q$\, are
fixed by the net baryon
number $B$\,, strangeness $S$ and electric charge $Q$\, of the system.

In heavy-ion collisions at high bombarding energies the partonic matter
created is characterized by $B\simeq 0$
and $S\simeq 0$, therefore $\mub_f\simeq\mu_f\simeq 0$\,.
On the other hand, strangelets are finite droplets with nonzero $B$ and
$S$\,. Using \re{ches} one can see that in this case
$\mu_B,\,\mu_S\neq 0$\, and, therefore, the inequality $\mu_s
>\mu_{u,\,d}$ should hold. This conclusion shows that strong
fluctuations are needed for strangelet formation in high energy nuclear
collisions.

If strangelet life times are long enough, the equilibrium with
respect to weak processes: \bel{weak} s\to
u+e^-\hspace*{-0.5ex}+\overline{\nu}_e,\,\,
u+e^-\hspace*{-0.5ex}\to s+\nu_e,\,\, \,s+u\leftrightarrow u+d,\,
\ee may be also achieved. Assuming that
$\mu_e=\mu_{\overline{\nu}}=0$ one arrives at the following
conditions \bel{wequ} \mu_u=\mu_d=\mu_s\,. \ee As will be shown
below, constituent masses $m_s$ of strange quarks in
(mechanically) stable strangelets exceed chemical potentials of
$u,d$ quarks. At $\mu_s>m_s$ and \mbox{$T\to 0$}\,\,the $\beta$
equi\-librium conditions (\ref{wequ}) can be realized only after
all $s$ quarks decay.

Using conditions (\ref{eqcp}) in the limit of zero temperatures one can see
that the density of antiquarks vanishes in baryon--rich chemically
equilibrated matter,\,\,$\rhob\to 0$\,. In the mean--field approximation
the reduced chemical potential of quarks with flavour $f$ coincides
with their Fermi energy:
\bel{efer}
\mu_{\mbs{Rf}}=\sqrt{m_f^2+p_{\mbs{Ff}}^2}\,,
\ee
where
\bel{fmom}
p_{\mbs{Ff}}=\left(\frac{6\pi^2\rho_f}{\nu}\right)^{1/3}
\ee
is the corresponding Fermi momentum.

Within the NJL model the energy density and pressure of matter as
well as the quark condensates $\rho_{\mbs{Sf}}$ contain divergent terms
originating from the negative energy levels of the Dirac sea. As noted
above, these terms are regularized by introducing the 3--momentum cutoff
\mbox{$\theta\,(\Lambda -|\bm{p}|)$}\,.
Then the scalar density is expressed as
\bel{cdens}
\rho_{\mbs{Sf}}=\nu\intp\frac{m_f}{E_{\mbs{\bm{p}f}}}
\left[\,\ocnf+\ocnbf-
\theta\,(\Lambda-p)\,\right]\,.
\ee

The energy density corresponding to
the Lagrangian (\ref{lagrm}) can be written~\cite{Mis00} as
\bel{enden}
e=e_K+e_D+e_S+e_V+e_{FM}+e_0\,.
\ee
This expression includes:\\
the ``kinetic'' term
\bel{ekin}
e_K=\nu\sum_f\intp\,E_{\mbs{\bm{p}f}}\left(\ocnf+
\ocnbf\right)\,,
\ee
the ``Dirac sea'' term
\bel{pdir}
e_D=-\nu\sum_f\intp\,E_{\mbs{\bm{p}f}}\,\theta\,(\Lambda -p)\,,
\ee
the scalar interaction term
\bel{pscal}
e_S=\frac{\ds G_S}{\ds 2}\sum_f\rho_{\mbs{Sf}}^{\,2}\,,
\ee
the vector interaction term
\bel{pvect}
e_V=\frac{\ds G_V}{\ds 2}\sum_f\rho_{\mbs{Vf}}^{\,2}
\ee
and the flavour mixing term
\bel{pfmix}
e_{FM}=-4K\prod_f\rho_{\mbs{Sf}}\,.
\ee
A constant $e_0$ is introduced in Eqs. (\ref{enden})
in order to set the energy density of the
physical vacuum equal to zero. This constant can be expressed through
the vacuum values of constituent masses, $m_f^{\rm vac}$, and quark
condensates, $\rho_{\mbs{Sf}}^{\rm vac}$. These values are obtained by
selfconsistently solving the gap equations (\ref{gape}) in vacuum, i.e.
at $\ocnf=\ocnbf=0$\,.

Explicit analytic formulae for the energy density and the gap
equations may be obtained in the case of zero temperature.
At $T\to 0$ one has~\cite{Mis99}
\bel{zer1}
e_K+e_D=\frac{\displaystyle\nu}{8\,\pi^2}\,\sum\limits_f
\left[\,p_{\mbs{Ff}}^4\Psi\left(\frac{\ds m_f}{\ds p_{\mbs{Ff}}}\right)
- \Lambda^4\Psi\left(\frac{\ds m_f}{\ds\Lambda}\right)\right]\,,
\ee
\bel{zer2}
\rho_{\mbs{Sf}}=\frac{\nu}{8\pi^2}\sum\limits_f
\left[\,p_{\mbs{Ff}}^{\,3}\,\Psi\,'\left(\frac{\ds m_f}
{\ds\,p_{\mbs{Ff}}}\right)
-\Lambda^3\Psi\,'\left(\frac{\ds m_f}{\ds\Lambda}\right)\right]\,,
\vspace*{1mm}
\ee
where
\bel{psid} \Psi(x)\equiv 4\int\limits_0^1 dt\,t^{\,2}\,\sqrt{t^{\,2}+x^2}
=\left(1+\frac{x^2}{2}\right)\sqrt{1+x^2}
-\frac{x^4}{2}\ln{\,\frac{1+\sqrt{1+x^2}}{x}}\,.
\ee

For a system with independent chemical potentials for quarks ($\mu_f$) and
antiquarks ($\mub_f$) one can use the thermodynamic identity
for the pressure of \qq matter
\bel{tiden}
P=\sum_f(\mu_f\rho_f+\mub_f\rhob_f)-e+sT\,,
\ee
where $s$ is the entropy density
\bel{entr}
s=-\nu\,\sum_f\intp\left[\,\ocnf\ln{\ocnf}+(1-\ocnf)\ln{(1-\ocnf)}
+ \ocnf\to\ocnbf\right]\,.
\ee

As discussed in Ref.~\cite{Mis99}, bound states of \qq matter
and first order phase transitions in \qq matter are possible
if its equation of state  $P=P(\mu_f,\mub_f,T)$ contains
regions with negative pressure or isothermal compressibility,
respectively. At $T=0$ the state of mechanical equilibrium
with vacuum ($P=0$) corresponds to minimum of the energy per particle,
\mbox{$\epsilon=E/\sum_f (N_f+N_{\bar{f}})=
e/\sum_f (\rho_f+\bar{\rho}_f)$}\,,
where $N_{f(\bar{f})}$ is the number of quarks (antiquarks) of flavour
$f$\,.
In the case of pure quark matter $N_{\bar{f}}=0$, its baryon number
$B=\sum_f N_f/3$ and its energy per baryon $E/B=3\epsilon$\,.

To characterize the flavour composition we introduce the
strangeness fraction parameter
\bel{rs}
r_s\equiv\frac{\ds S}{3B}=\frac{\ds\rhob_s-\rho_s}{\ds 3\rho_B}\,.
\ee
Below only the isospin--symmetric mixtures where $N_u=N_d$ and
$N_{\bar{u}}=N_{\bar{d}}$ will be considered. It can be
shown~\cite{Mis00} that in the dilute limit, when all single flavour
densities $\rho_f, \rhob_f$ are small, $\epsilon$ tends to the sum of
the constituent quark and antiquark masses in vacuum, weighted
according to $r_s$\,:
\bel{vmrs}
m_q^{\rm vac}(r_s)=
(1-r_s)\,m_u^{\rm vac}+r_s\,m_s^{\rm vac}.
\ee
In the case of chemically equilibrated matter at $T=0$
and fixed $r_s$ one has
\bel{vmrs1}
\epsilon (\rho_B\to 0, r_s)=m_q^{\rm vac}(r_s)
\ee
A bound multiparticle state exists if there is a nontrivial
minimum of $\epsilon$ as function of $\rho_B$ and
the binding energy (BE) per baryon is positive:
\bel{ben}
{\rm BE}=3\,\left[m_q^{\rm vac}(r_s)-\epsilon_{\rm min}(r_s)\right]>0\,.
\ee

\section{Strange quark matter at zero temperature}

Let us consider first quark matter with nonzero net baryon density at
$T=0$. In the chemically equilibrated system the density of valence
antiquarks will be zero for each flavour ($\rhob_{f}=0$). Fig.~1 shows
the energy per quark as a function of baryon density,
$\rho_B=\frac{1}{3}\sum\limits_f\rho_f$\,.  Different curves correspond
to different $r_s$, which in this case is the relative concentration of
strange quarks. In accordance with~\re{vmrs1}, at $\rho_B\to 0$ the
energy per quark tends to the corresponding vacuum mass. With growing
density both the attractive scalar and repulsive vector interactions
contribute to $\epsilon$ (see the discussion of this question in
Ref.~\cite{Mis00}).

It is interesting that at $r_s\leq 0.7$ the attractive interaction is
strong enough to produce a nontrivial local minimum at a finite
$\rho_B$. In the pure $u, d$ matter ($r_s=0$) this minimum is unbound
by about 20 MeV as compared to the vacuum masses of $u$ and $d$ quarks.
On the other hand, it is located at a baryon density of about
$1.8\,\rho_0$, which is surprisingly close to the saturation density of
normal nuclear matter.  Of course, the location of this minimum depends
on the model parameters. Nevertheless, one can speculate that
nucleon--like 3--quark correlations, not considered in the mean--field
approach, will turn this state into the correct nuclear ground state.

When $r_s$ grows from 0 to about 0.4, the local minimum is getting
more pronounced and the corresponding baryon density increases to about
$3.2\,\rho_0$. At larger $r_s$, the minimum again becomes more shallow
and disappears completely at $r_s\simeq 0.7$. At \mbox{$0.2<r_s<0.6$}
the minima correspond to the true bound states, i.e. the energy per
quark is lower than the respective vacuum mass. But in all cases these
bound states are rather shallow: even the most strongly  bound state at
$r_s\simeq 0.4$ is bound only by about 15~MeV per
baryon. Nevertheless, the appearance of local minima signifies the
possibility for finite droplets to be in mechanical equilibrium with
the vacuum at $P=0$. It is natural to identify such droplets with
strangelets, which are hypothetical objects made of light and strange
quarks~\cite{Wit84,Far84,Gre87,Gil93,Schaf98,Wil99}.

It should be emphasized here that $\beta$--equilibrium is not
required in the  present approach (see the discussion below).
That is why our most bound strangelets are predicted to be more
rich in strange quarks ($r_s>1/3$) than in the approaches assuming
$\beta$--equilibrium~\cite{Wit84,Far84,Gil93}, which give
$r_s<1/3$. As a result, these strangelets will be negatively
{charged\footnote{Negatively--charged strangelets have been also
considered in Refs.~\cite{Gre87,Schaf98}.}}. Indeed, the ratio of
the charge $Q$ to the baryon number $B$ is expressed through $r_s$
as
\bel{q}
\frac{Q}{B}=\frac{2}{3}\frac{\rho_u}{\rho_B}-\frac{1}{3}
\frac{\rho_d}{\rho_B}
-\frac{1}{3}\frac{\rho_s}{\rho_B}=\frac{1}{2}(1-3r_s).
\ee
For $r_s\simeq 0.4$ this gives $Q/B\simeq -0.1$. In light of recent
discussions (see e.g.~Ref.~\cite{Wil99}) concerning possible
dangerous scenarios of the negatively--charged strangelet production at
RHIC, we should emphasize that the strangelets predicted here are not
absolutely {bound\footnote{The analogous conclusion has been made in
Ref.~\cite{Bub99}.}}, i.e. their energy per baryon is higher than that
for the normal nuclear matter. Hence, the spontaneous conversion
of normal nuclear matter to strange quark matter is energetically
not possible.

Fig.~2 shows the constituent masses of $u$ and $s$ quarks as functions
of baryon density. The dropping masses manifest a clear tendency
to the restoration of chiral symmetry at high densities.
The dots indicate the masses at the local minima in the
respective energies per baryon shown in Fig.~1.
Note that the stronger is reduction of constituent
masses the deeper are the corresponding bound states. For the
metastable state at $r_s=0$, which is a candidate for the nuclear
ground state, the masses of $u$ and $s$ quarks are equal to 0.3 and 0.9
of their vacuum values respectively. For the most bound state at
$r_s\simeq 0.4$ the respective mass ratios are reduced to 0.15 and 0.6.

The behaviour of the $u$ and $s$ chemical potentials is shown in
Fig.~3. At $r_s\neq 0, 1$, due to the contribution of the vector
interaction (see Eqs.~(\ref{rce1}), (\ref{efer})), $\mu_u$ and
$\mu_s$ grow practically linearly at large baryon densities. In
accordance with a previous discussion, one can see that the
conditions $\mu_s>\mu_{u,d}$ hold at baryon densities
corresponding to bound states of strange matter.

The properties of the multiparticle bound states are summarized in
Figs.~4--5. Fig.~4 shows the binding energy per baryon, \re{ben}.
The maximum binding, about 15~MeV, is realized at $r_s\simeq 0.4$.  One
should bear in mind, that in the case of baryon--rich matter local
minima of $\epsilon$ result from a strong cancellation between the
attractive scalar and repulsive vector interactions.  Therefore, it is
very sensitive to their relative strengths. The results presented above
are obtained for $G_V=0.5\,G_S$.  For comparison in Figs.~4 and 5 we
also present the model predictions for $G_V=0$. In this case the
maximum binding energy increases to about 90~MeV per baryon and the
corresponding $r_s$ value shifts to about 0.6. It is interesting to
note that for \mbox{$G_V=0$} the bound state appears even in the pure
$u,d$ matter. The corresponding binding energy is about 20 MeV per
baryon.

The dots in Fig.~5 indicate the positions of some conventional baryons.
By inspecting the figure, one can make a few interesting observations.
First, conventional baryons are more bound than
strangelets even at $G_V=0$\,. This is an indication that baryon--like
3--quark correlations might be indeed very important in the
baryon--rich quark matter. Second, the bound state energies grow
monotonously with $r_s$\,.

\section{Quark matter at finite temperatures}

In this section we study properties of deconfined matter at finite
temperatures. The calculations for this case can be done by using
general formulas of Sect.~II with the quark and antiquark occupation
numbers given by Eqs.~(\ref{ocnq})--(\ref{ocnqb}). Unless stated
otherwise, the results given below correspond to the ratio
$G_V/G_S=0.5$\,.

Figs.~6(a) represents the pressure isotherms for the case of
zero net strangeness (\mbox{$r_s=0$}), which is appropriate for fast
processes where net strangeness is conserved, e.g. in relativistic
nuclear collisions.  One can see a region of spinodal instability,
\mbox{$\partial_\rho P<0$}, which is characteristic for a first order
phase transition.  The corresponding critical temperature is about 35
MeV. The dashed line (binodal) shows the boundary of mixed phase
states. Bound (zero--pressure) states exist only at temperatures below
15~MeV.

Generally, the equation of state of the chemically equilibrated quark
matter is characterized by two quantities: net baryon charge, $B$, and
net strangeness, $S$. Therefore, it is interesting to study thermal
properties of this matter at $S\neq 0$\,. Such states can be reached in
neutron stars. They can also be realized via the distillation mechanism
accompanying a QCD phase transition in heavy--ion collisions
\cite{Car91}. Fig.~6(b) shows the pressure isotherms for
\mbox{$r_s=0.4$}. As discussed above, this case corresponds to most
bound strange matter at $T=0$\,. At this value of $r_s$ bound states
exist at $T<30$\,\,MeV. The dashed line in Fig.~6(b) again shows
the boundary of the two--phase region. It is obtained by the solution
of the Gibbs conditions
\mbox{$P^{(1)}=P^{(2)}$}, \mbox{$\mu_u^{(1)}=\mu_u^{(2)}$},
\mbox{$\mu_s^{(1)}=\mu_s^{(2)}$} where indices 1 and 2 denote two
coexisting phases. As compared to the case $r_s=0$ the chiral phase
transition takes place in a wider region of $T$ and $\rho_B$\,.
Applying the Gibbs conditions, one can see that for $r_s=0$ the
equilibrium pressure isotherms are constant in the mixed phase region
(the Maxwell construction). But this is not the case for $r_s=0.4$\,.
In accordance with general conclusions of Ref.~\cite{Gle92}, in the case
of two conserving charges (baryon number and strangeness) the Maxwell
construction is modified in such a way that the equilibrium pressure
at fixed $T$ increases with $\rho_B$ in the mixed--phase domain.
However, this increase is small (several percents) and hardly visible in
Fig.~6(b). For example, for $T=30$ MeV the equilibrium pressure
changes from  5.36 to 5.44 MeV/fm$^3$. It is interesting, that
local values of $r_s$ in the coexisting phases are slightly different
($r_s^{(1)}<0.4<r_s^{(2)}$) and only the global strangeness ratio is
fixed ($S/3B=0.4$).

Fig.~7 shows the critical temperatures for the existence of phase
transitions and bound states in the equilibrated strange matter as
functions of $r_s$\,. One can see that both temperatures first grow
with $r_s$ and then drop to zero at $r_s\sim 0.8$. The maximal values
of respectively 50 MeV and 30 MeV are realized at some intermediate
$r_s\sim 0.4$\,. As demonstrated earlier in Fig.~4, this value of $r_s$
corresponds to the most bound state of strange quark matter at $T=0$.
So we see an obvious correlation: the deeper is a bound state at $T=0$
the stronger is a phase transition at finite temperatures.

It should be emphasized here again that the thermal properties of
asymmetric baryon--rich quark matter are very sensitive to the relative
strength of scalar and vector interactions~\cite{Mis00}. If we take
$G_V=0$, as in most calculations in the literature, the corresponding
critical temperature at $r_s=0$ increases to about 70 MeV. On the other hand,
if one takes $G_V=0.65\,G_S$, the zero--pressure states disappear
completely.  The calculation shows that there is no phase transition at
$G_V>0.71\,G_S$\,. It is interesting to note that in all cases this
first order phase transition occupies the region of densities around
the normal nuclear density $\rho_0$\,.

A more detailed information about the first order (chiral) phase
transition predicted by our model is given in Figs.~8 where the
critical temperature $T_c$ and baryon chemical potential
$\mu_B=\mu_u+2\mu_d=3\mu_u$ are shown for different values of $r_s$\,.
Again one can see that maximal $T_c\simeq 50$\,MeV corresponds to
$r_s\simeq 0.4$\,.

Two phases coexisting in the mixed-phase domain of the chiral phase
transition are characterized by different values of constituent quark
masses. If \qq matter evolves from a high to a low density state,
crossing the two--phase domain, the transition from chirally restored
states (with low quark masses $m_f\sim m_{0f}$) to the chirally broken
phase (where \mbox{$m_f\sim m_f^{\rm vac}$}) takes place. Therefore,
a part of the internal energy should be transformed into the rest mass.
If the total
entropy \footnote{This quantity should not be confused with the net
srangeness introduced in Sec.~II.} $S$ and the effective number of
degrees of freedom are conserved, the temperature will decrease
during this transition. This is clearly seen in Fig.~9, where
isentropes with entropy per baryon $S/B=2$ and $S/B=5$ are shown for the
case $r_s=0$\,. The analogous effect has been predicted within the
linear $\sigma$ model in Ref.~\cite{Sca00}. The drop of temperature as a
result of this cooling mechanism may serve as an observable
signature of the chiral phase transition in nuclear collisions. The
above behaviour is qualitatively different from the one expected in the
deconfinement transition where the number of degrees of freedom is
smaller in the dilute (hadronic) phase and the temperature may increase
during hadronization~\cite{Hal98}.

These results demonstrate that the chiral phase transition is rather
similar to a liquid--gas phase transition in normal nuclear matter.
The critical temperature and baryon density in the present case
($T_c\sim 30$ MeV, $\rho_{Bc}\sim\rho_0$) are not so far from the
values predicted by the conventional nuclear models~\cite{Goo85}
($T_c\sim 20$ MeV, $\rho_{Bc}\sim 0.5\,\rho_0$)\,. One may expect that
in a more realistic approach, taking into account nucleonic
correlations, the chiral transition may turn into the ordinary
``liquid--gas'' phase transition. If this would be the case, one
should be doubtful about the possibility of any other QCD phase
transition of the liquid--gas type at a higher baryon density. At least
only one phase transition of this type is predicted within the NJL
model.

\section{Discussion of decay modes}

Let us discuss briefly possible decay channels of bound states of
quark matter described above. In strange quark matter (without
antiquarks), flavour conversion is only possible through weak
decays. As follows from Fig.~3, at densities corresponding to zero
pressure, the condition \mbox{$\mu_s>\mu_u$} holds. This means
that weak processes of the types \mbox{$s\to
u+e^{-}+\overline{\nu}_e$\,,}\,\,\mbox{$s+u\to u+d$} are allowed.
Since there is no local barrier at any $r_s$ (see Fig.~5), in a
system produced initially at some $r_s\neq 0$, all strange quarks
will eventually  be converted into light $u, d$ quarks.
Schematicly this conversion process is shown in Figs.~9(a) where
the initial ($r_s\ne 0$) and final ($r_s=0$) states corresponds to
the left and right diagram, respectively.

This picture is very different as compared to the one based on the MIT
bag model. Because the quark masses are kept constant ($m_f=m_{0f}$),
the condition $\mu_s>m_s$ will be first satisfied at a relatively low
baryon density $\propto m_{0s}^3$. At higher densities a certain
fraction of $s$ quarks will always be present in a
$\beta$--equilibrated matter (see Fig.~9(b)).

In contrast, in the NJL model the $s$ quark mass is a function of both
baryon density and strangeness content. As one can see from Fig.~9(a),
at any given $r_s$ the condition $\mu_{u, d}=\mu_s$ can be satisfied
only at sufficiently high baryon densities which correspond to positive
pressure. On the other hand, at the points of zero pressure we always
have $\mu_{u, d}<\mu_s$. Therefore, weak decays will proceed until a
system reaches $r_s=0$ (see right picture in Fig.~9(a)).

The life times of strangelets at $T=0$ can be roughly estimated by analogy to
the neutron decay.
The matrix element of s-quark $\beta$ decay is proportional to
$(\Delta E)^{5/2}\sin{\theta_c}$  where $\Delta E$ is the energy gain
in the reaction and $\theta_c$ is the Cabibbo angle
($\sin{\theta_c\simeq 0.22}$~\cite{PDG}). As follows from our
calculations (see Fig.~3), the energy  gain $\Delta E=\mu_s-\mu_u$
depends on $r_s$, ranging from about 200 MeV at $r_s=0.4$ to 100 MeV at
$r_s=0$\,.  By scaling with corresponding quantities for the neutron
decay, one can express the life time of a strangelet as
\bel{ltes}
\tau\sim\tau_n\left(\frac{\Delta m}{\Delta E}\right)^5
\sin^{-2}{\theta_c},
\ee
where $\Delta m=m_n-m_p\simeq 1.2$ MeV and
$\tau_n\simeq 880\,s$ is the life time of neutron~\cite{PDG}.  This
estimate gives $\tau \sim 10^{-9}\div 10^{-7}$ depending on $r_s$.  It
is not surprising that this is close to the life times of charged pions
and conventional hyperons.

\section{Conclusions}

By using the NJL model, we have investigated the equation of state of
chemically equilibrated deconfined quark matter at various temperatures,
baryon densities and strangeness contents. The model predicts the
existence of loosely bound, negatively--charged strangelets with
maximal binding energies of about 15 MeV per baryon at $r_s\sim 0.4$\,.
Similarly to Ref.~\cite{Bub99}, no absolutely stable strange quark
matter has been found. The estimated life--times of these states may be as
long as $10^{-7}$ s. It is shown that properties of baryon--rich quark
matter are very sensitive to the relative magnitude of the vector and
scalar interactions. At the standard values of vector and scalar
couplings, $G_V/G_S=0.5$, the metastable bound states of chemically
equilibrated matter exist at $T<15$ MeV, while at $G_V=0$ this
temperature increases to 40 MeV.

The calculations reveal the first order chiral phase transition at finite
baryon densities and moderate temperatures. In the case of zero net
strangeness ($r_s=0)$ the critical
temperature is in the range of 30 MeV and the critical baryon density
is around~$\rho_0$. We believe that this phase transition is
reminiscent of the ordinary liquid--gas phase transition in nuclear
matter. We have found that the maximal critical temperature $T_c\simeq
50$ MeV is reached for the ratio of net strangeness to baryon charge
$S/B\simeq 1.2$\,. The model predicts a strong cooling of matter during the
phase transition due to generation of the constituent mass.

We hope that these results will generate a certain optimism in searching for
unusual states of matter in relativistic heavy-ion collisions.

\section*{Acknowledgments}

This work has been supported by the RFBR Grant No. 00--15--96590,
the Alexander von Humboldt Foundation, the Graduiertenkolleg ``Experimentelle
und Theoretische Schwerionenphysik'', GSI, BMBF and DFG.

\newpage
\begin{figure}
\centerline{\psfig{figure=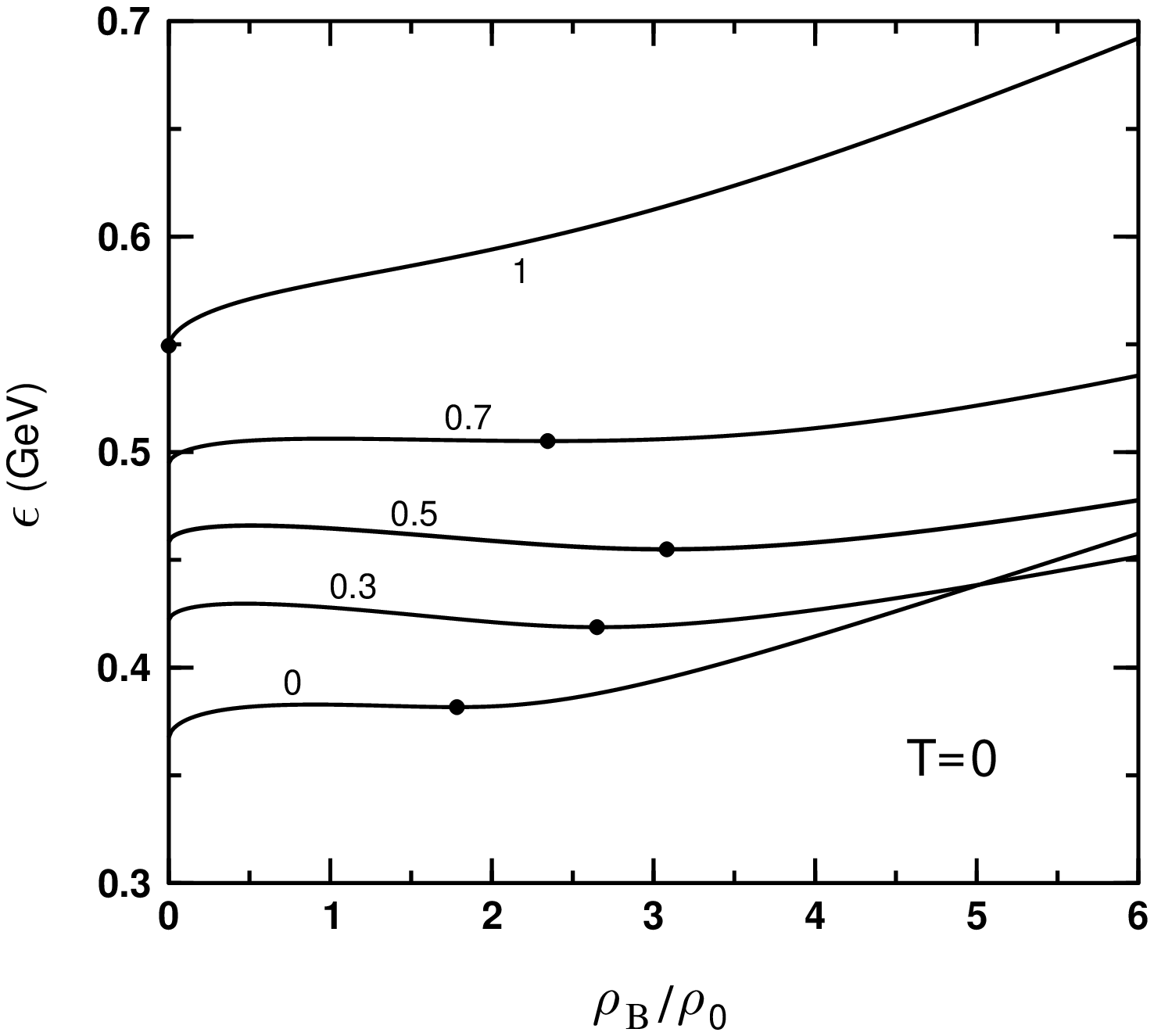,height=15cm}}
\vspace*{1.5cm}
\caption
{Energy per particle $\epsilon$ in pure quark matter at zero
temperature as function of  baryon density at different values of
strangeness fraction $r_s$.  Points indicate local minima of
$\epsilon$.  $\rho_0=0.17\,{\rm fm}^{-3}$ is normal nuclear density.}
\end{figure}
\begin{figure}
\centerline{\psfig{figure=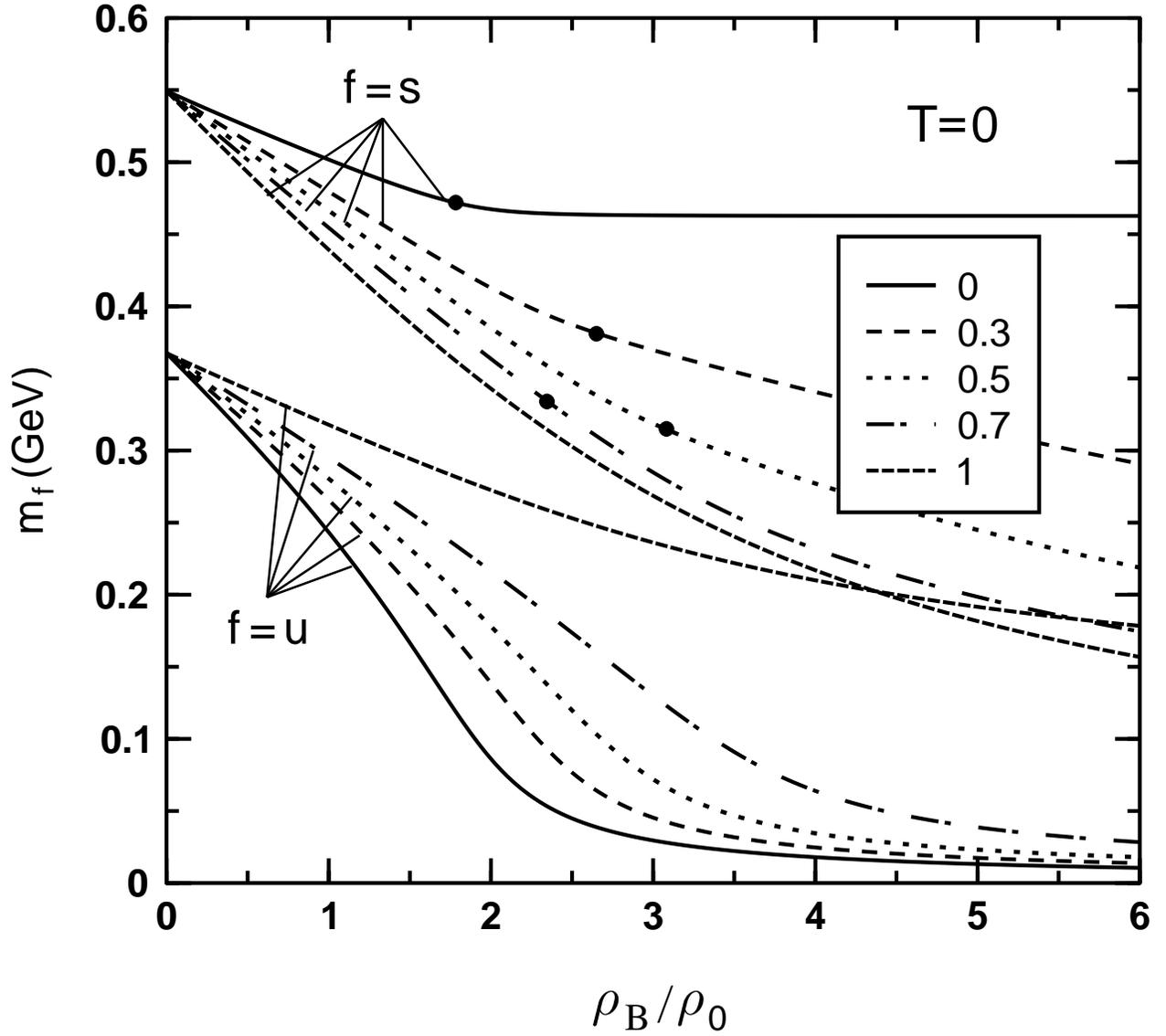,height=15cm}}
\vspace*{1.5cm}
\caption
{Constituent masses of $u$ and $s$ quarks vs baryon density.
Figures in the box show values of strangeness fraction $r_s$. Dots
correspond to minima of energy per particle at given $r_s$.}
\end{figure}
\begin{figure}
\centerline{\psfig{figure=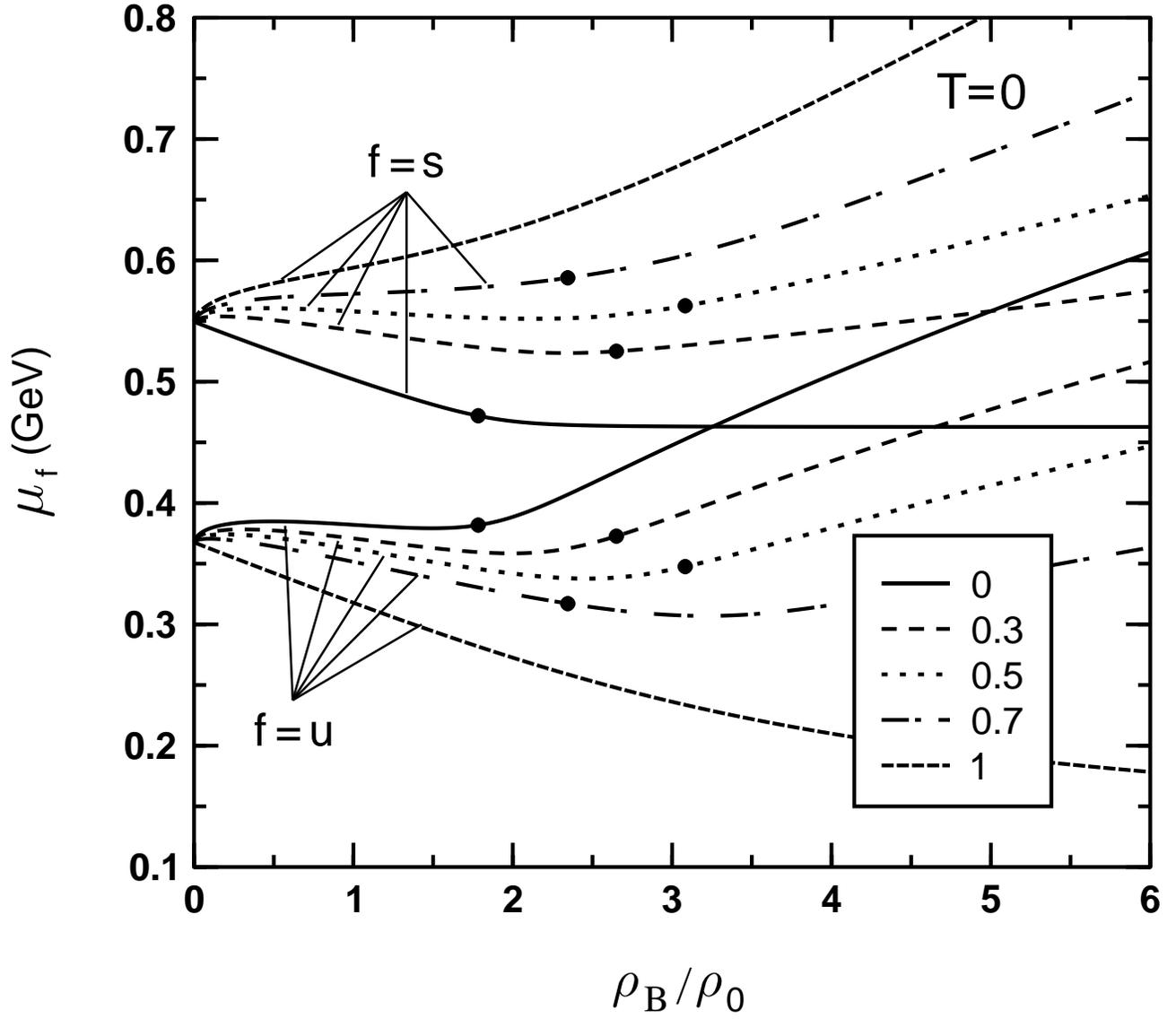,height=15cm}}
\vspace*{1.5cm}
\caption
{The same as in Fig. 3, but for chemical potentials of $u$ and
$s$ quarks.}
\end{figure}
\begin{figure}
\centerline{\psfig{figure=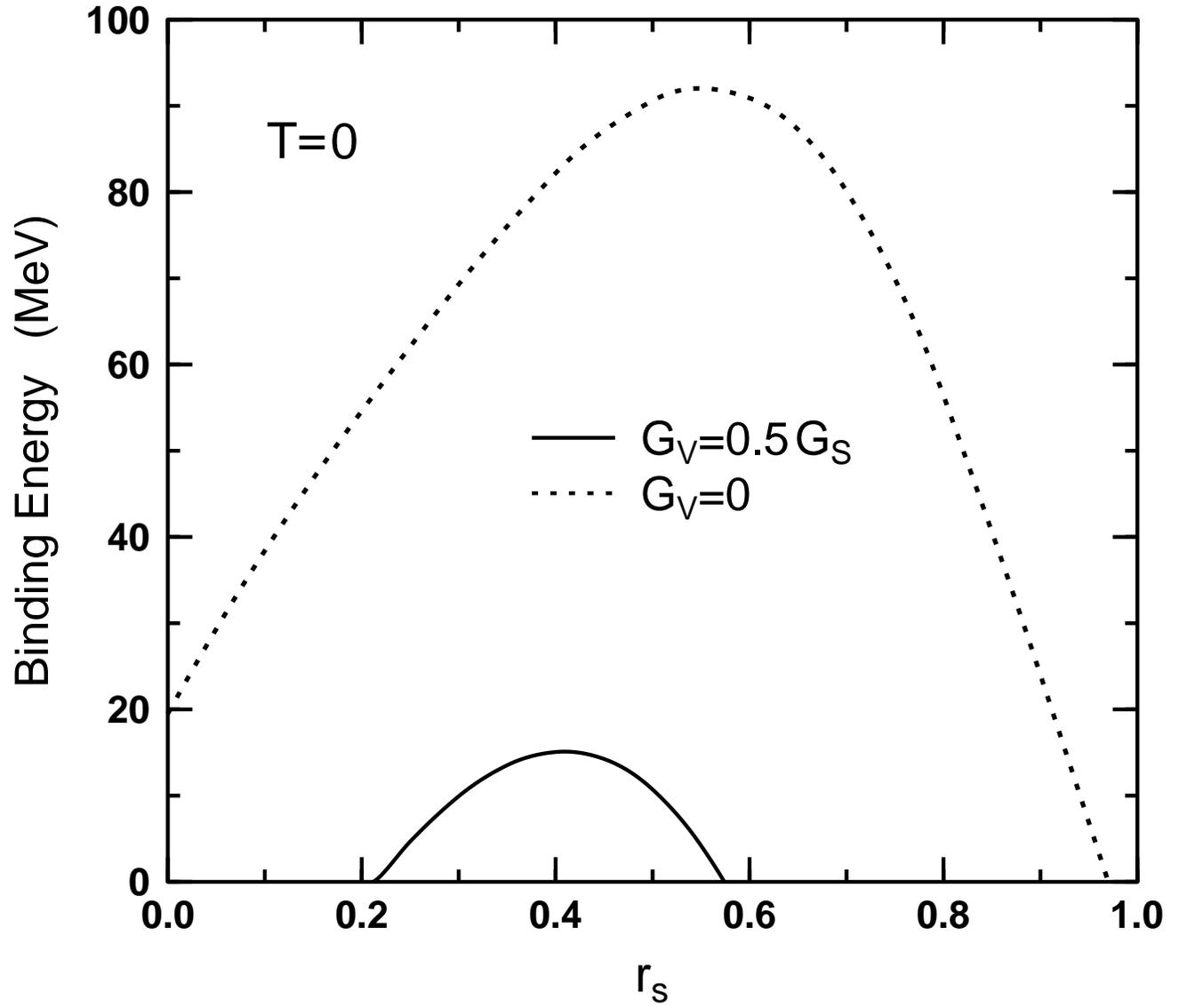,height=15cm}}
\vspace*{1.5cm}
\caption
{Binding energies per baryon in  pure quark matter as functions
of strangeness fraction~$r_s$.  Dotted line shows the results of
calculations when the vector interaction is switched off~($G_V=0$).}
\end{figure}
\begin{figure}
\centerline{\psfig{figure=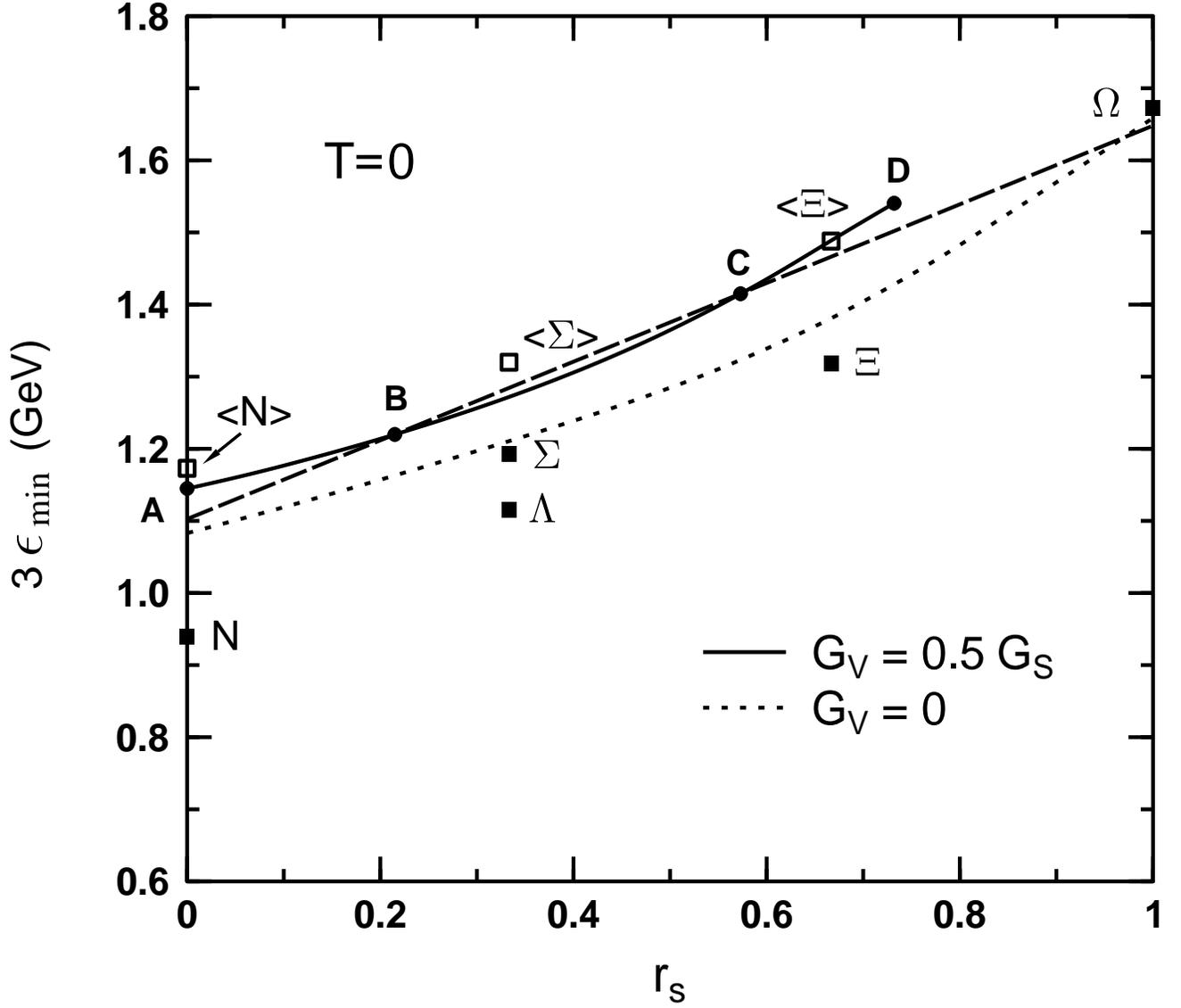,height=15cm}}
\vspace*{1.5cm}
\caption
{Minimal energies per baryon in pure quark matter as functions
of strangeness fraction~$r_s$. Dashed line shows the same energy in the
limit of zero particle densities, Eq.~(\ref{vmrs1}). Different parts of
the solid line correspond to metastable (AB and CD) or bound (BC)
states.  Filled squares show masses of lightest baryons. Open squares
represent the spin--isospin averaged masses of these baryons, e.g.
\mbox{$<N>=[m_{\,N}+4\,m_{\,\Delta(1232)}]/5$} (for details
see Ref.~[19]). Dotted line in the lower plot shows the results
in the limit $G_V\to 0$.}
\end{figure}
\begin{figure}
\centerline{\psfig{figure=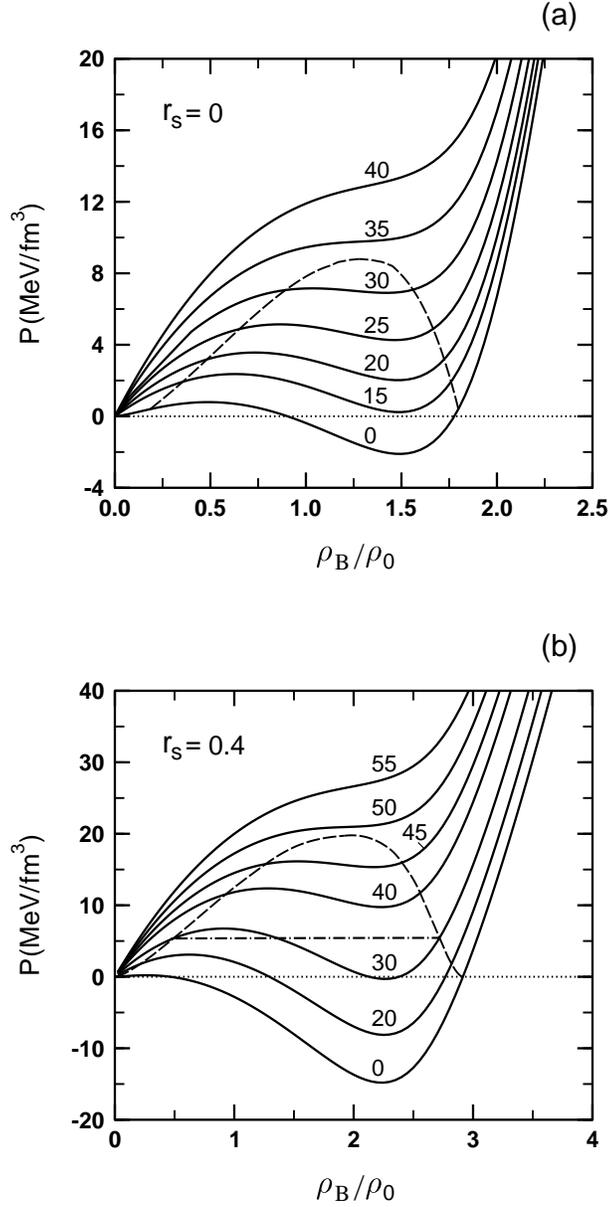,height=16cm}}
\vspace*{1.5cm}
\caption
{Pressure isotherms for chemically equilibrated quark matter
with \mbox{$r_s=0$} (upper part) and \mbox{$r_s=0.4$} (lower part).
Temperatures are given in MeV near the corresponding curves. Boundaries
of spinodal regions are shown by the dashed lines.
The dashed--dotted line in lower plot shows the equilibrium pressure
in the mixed phase--domain at $T=30$ MeV.}
\end{figure}
\begin{figure}
\centerline{\psfig{figure=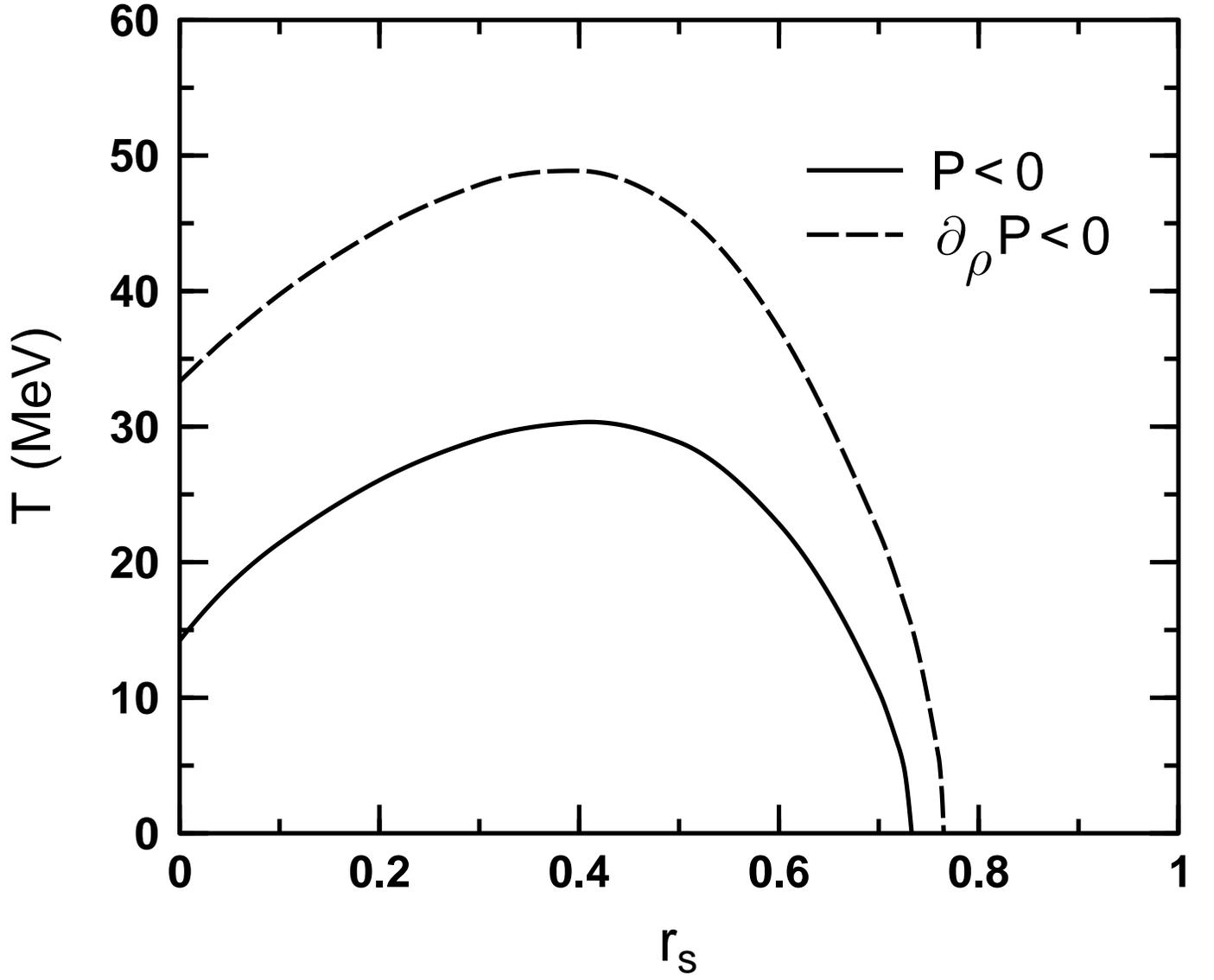,height=15cm}}
\vspace*{1.5cm}
\caption
{Critical temperatures for existence of bound states ($P<0$)
and phase transition ($\partial_\rho P<0$) in equilibrium quark matter
as functions of strangeness fraction.}
\end{figure}
\begin{figure}
\centerline{\psfig{figure=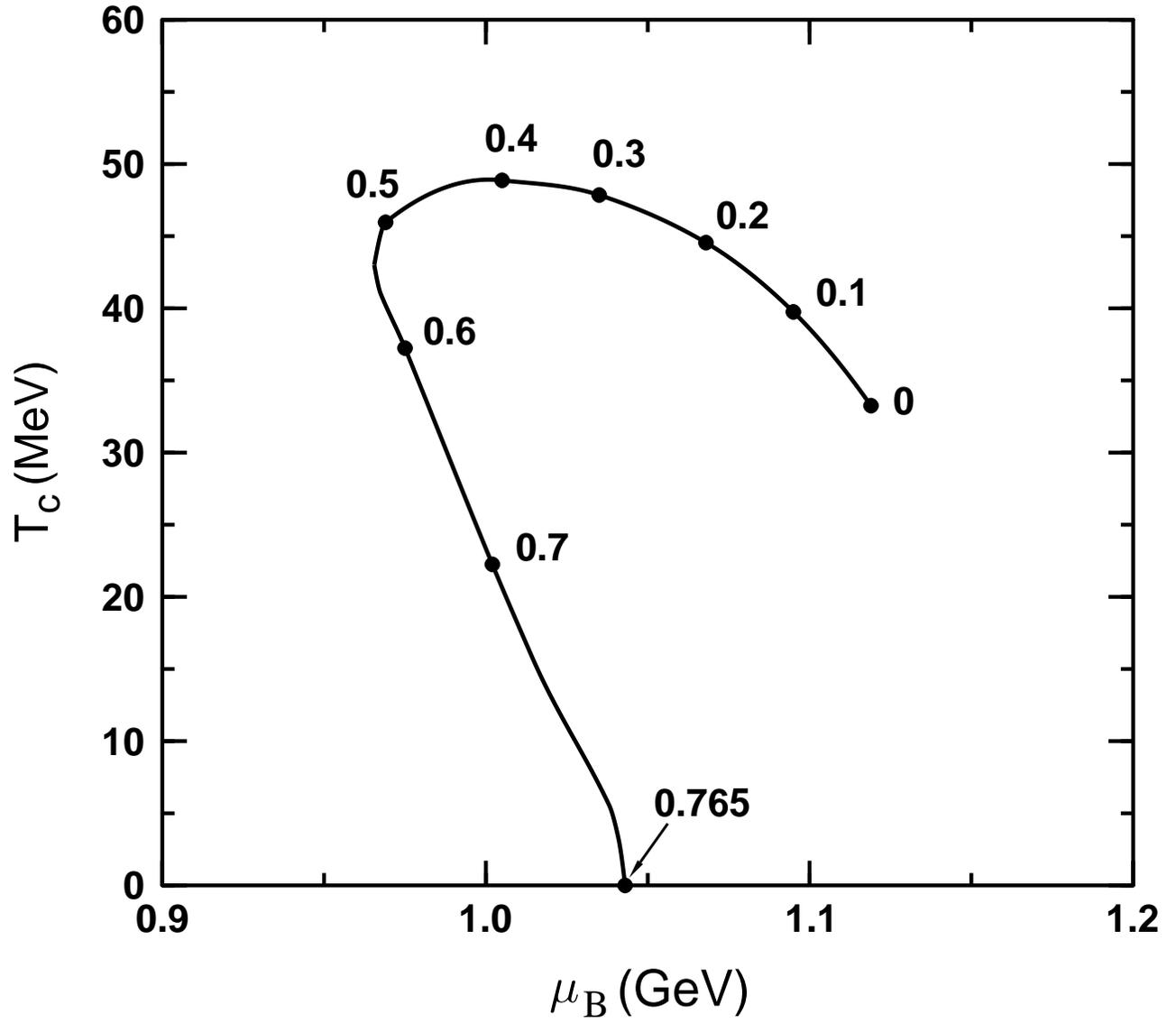,height=15cm}}
\vspace*{1.5cm}
\caption
{Critical temperatures and baryon chemical potentials in
equilibrated quark matter at fixed strangeness fractions $r_s$
(indicated by figures near corresponding points).}
\end{figure}
\begin{figure}
\centerline{\psfig{figure=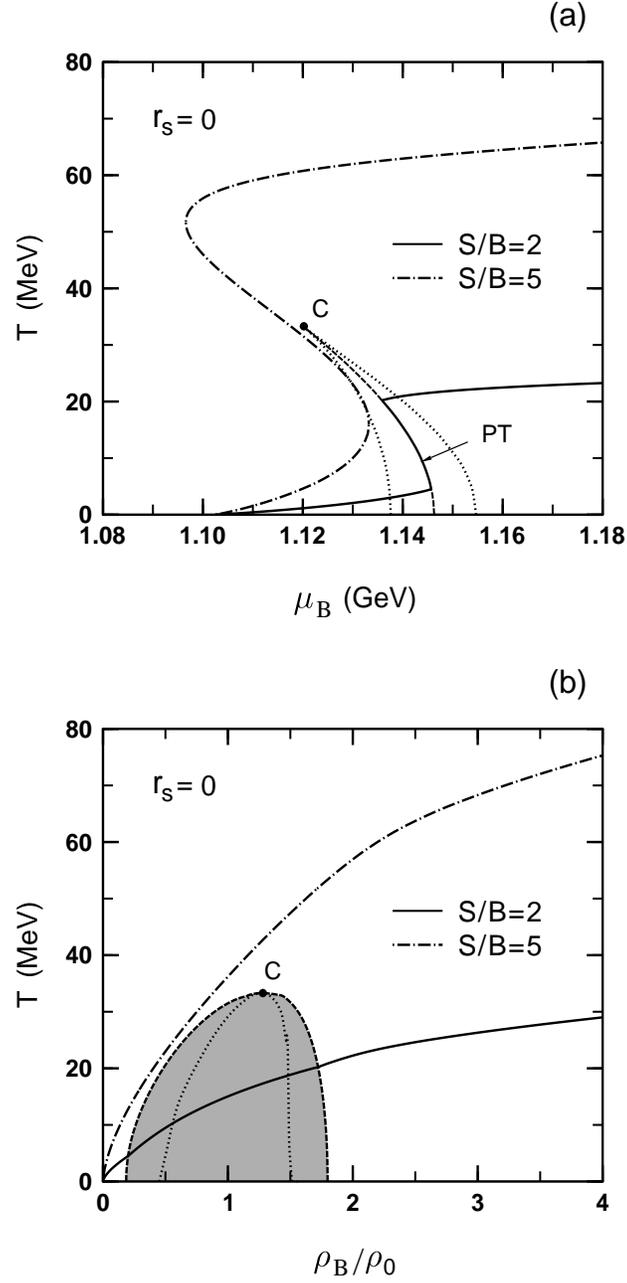,height=17cm}}
\vspace*{1.5cm}
\caption
{Isentropes $S/B=2$ (solid line) and $S/B=5$ (dashed--dotted
line) of quark matter with zero net strangeness. Lower (upper) part
corresponds to the $\mu_B-T$ ($\rho_B-T$) plane. Dotted and dashed
lines represent, respectively, spinodal and binodal boundaries of the
two--phase region (shown by shading in the lower plot). Point C marks
the critical point.}
\end{figure}
\begin{figure}
\centerline{\psfig{figure=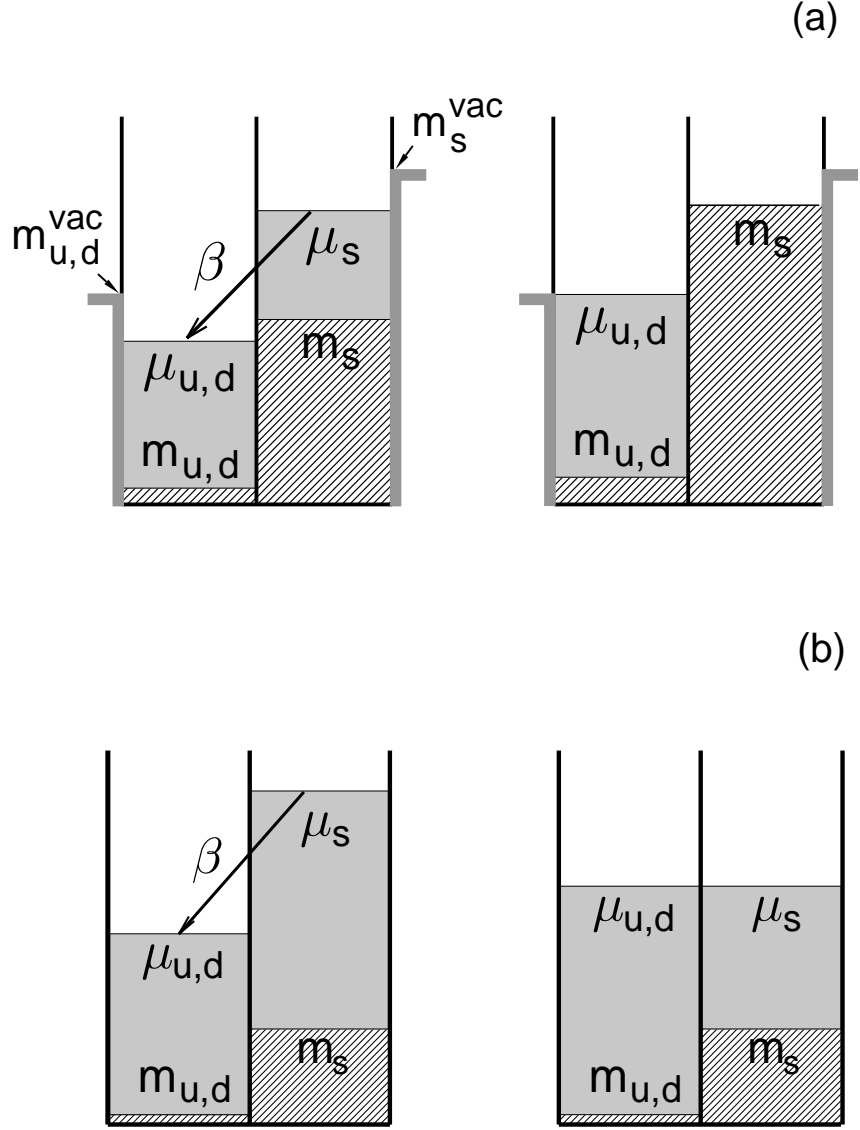,height=15cm}}
\vspace*{1.5cm}
\caption
{Schematic pictures of energy levels (shown by shading)
occupied by light and strange quarks in cold quark matter at different
strangeness fractions $r_s$. Left and right upper diagrams~(a) shows
the results for strange ($r_s=0.4$) and nonstrange (\mbox{$r_s=0$})
matter predicted within the NJL model. Low panel (b) represents the
same results within the MIT bag model~[19]. Hatched boxes in
upper and low panels shows, respectively, the constituent and bare
quark masses. Arrows show weak decay processes \mbox{$s\to
u+e+\overline{\nu}_e$} and \mbox{$s+u\to u+d$}.}
\end{figure}
\end{document}